# Metastable phase in the quantum Hall ferromagnet

Vincenzo Piazza[a*], Vittorio Pellegrini[a], Fabio Beltram[a], Werner Wegscheider[bc]

[a] *NEST-INFM and Scuola Normale Superiore, Pisa (Italy)*
[b] *Walter Schottky Institut, Technische Universität München, D-85748 Garching, (Germany)*
[c] *Institut für Experimentelle und Angewandte Physik, Universität Regensburg, D-93040 Regensburg, (Germany)*

Time-dependent capacitance measurements reveal an unstable phase of electrons in gallium arsenide quantum well that occurs when two Landau levels with opposite spin are brought close to degeneracy by applying a gate voltage. This phase emerges below a critical temperature and displays a peculiar non-equilibrium dynamical evolution. The relaxation dynamics is found to follow a stretched exponential behavior and correlates with hysteresis loops observed by sweeping the magnetic field. These experiments indicate that metastable randomly-distributed magnetic domains with peculiar excitations are involved in the relaxation process in a way that is equivalently tunable by a change in gate voltage or temperature.

## 1. Introduction

At low temperatures and under the application of high magnetic field (B), electronic interactions dominate over single-particle potentials thanks to the quantization of in-plane kinetic energy into macroscopically degenerate Landau levels. Magnetic-field coupling to electron spin further splits each Landau level into spin-up and spin-down states separated by the Zeeman energy. The filling factor $\nu = n_e 2\pi l_B^2$ (here $n_e$ is the surface electron density and $l_B = (\hbar c/eB)^{1/2}$ is the magnetic length, $\hbar$ is the Planck's constant) defines the number of Landau levels occupied by electrons. QH states are distinct phases of electrons that are observed at integer and some fractional values of $\nu$[1,2]. Fractional quantum Hall states, in particular, results from the creation of highly-correlated electron liquids driven by Coulomb repulsion[1].

---

[*] piazza@nest.sns.it, fax n.: +39 050 509417

The impact of Coulomb interaction in the integer regime is emphasized at $\nu = 1$. In this case exchange effects can lead to spontaneous ferromagnetism which, in fact, can persist even in the limit of zero Zeeman energy. The ground state in this configuration is an isotropic itinerant ferromagnet[3] with peculiar charge excitations[4,5]. It is this competition between direct and exchange terms of the Coulomb interaction on one hand, and single-particle potentials on the other that is able to drive the ground states into non-trivial broken-symmetry phases that are connected by quantum phase transitions. A remarkable manifestation of this interplay was first pointed out by Giuliani and Quinn[6]. These authors predicted a first-order phase transition between spin-polarized and spin-unpolarized quantum Hall states near crossing between different Landau levels in a tilted magnetic field configuration, a phenomenon that was successively investigated in several experiments[7,8]. The complexity of these broken-symmetry states can be further enlarged if new internal degrees of freedom are added to the system. This is the case often encountered in semiconductor heterostructures with two parallel coupled 2DEGs. Drawing an analogy between the layer-index degree of freedom and spin, the former is often described in terms of a *pseudospin* quantum number. The splitting between pseudospin up and down energy levels plays the role of the Zeeman energy and determines the ground-state properties. It can be tailored by adjusting layer separation and tunnelling strength by means of external perturbations. Electronic interactions can lead to pseudospin anisotropies[9] that in the simplest case occurs when two Landau levels with opposite spins and/or pseudospin become degenerate in the integer regime. These issues stimulated a large body of experimental as well as theoretical work that are unveiling some of the most intriguing phenomena in semiconductor many-body physics[10].

The magneto-transport signatures of Ising QH ferromagnets are hysteresis and the collapse of the QH state close to level crossing. It is now established that these features originate from the presence of a multi-domain structure that is induced by disorder below a critical temperature. These magnetic domains lead to dissipation paths for mobile carriers which are allowed to move from one edge of the sample to the other by scattering between domain walls[11,12]. Domain-wall excitation properties were only recently modelled[13,14]. It is expected that conformation and length of domain walls below a critical temperature $T_c$ lead to large energy barriers between adjacent domains. A metastable configuration of magnetic domains with slow evolution should then occur and be responsible for the hysteresis observed in up and down sweeps of $B$. Evidence of such a slow evolution was recently reported at the transition between states with opposite spin and/or pseudospin polarization in the fractional QH regime[15,16,17].

In this communication we report a set of experiments that show the existence of a metastable phase of magnetic domains of an Ising QH ferromagnet in which non-exponential relaxation times can be induced below a critical temperature and carefully controlled by changing the level alignment with the gate bias. In the studies here discussed the magneto-capacitance ($C_B$) between the 2DEGs and a top metal gate was measured as a function of time. Since experimental $C_B$ values are determined by the area carrying the edge current[18], these experiments represent a direct probe of the evolution of magnetic-domain configurations in the Ising QH ferromagnet. While these results improve our understanding of the peculiar low-energy excitation pattern of DWs in a QH ferromagnet, they also represent a new tool for the experimental study of the evolution of low-dimensional electron systems towards equilibrium.

## 2. Experiment

The sample studied in this work is a modulation-doped, GaAs/Al$_{0.25}$Ga$_{0.75}$As heterostructure containing a wide (60 nm) quantum well (QW). A "soft" barrier, originating from Coulomb interactions in the QW, separates the electronic system into two well-defined layers. Indium Ohmic contacts provide electrical access to the 2DEGs. After illumination, at equilibrium and at a temperature of 4.2 K the population of the top 2DEG is $3.9 \times 10^{11}$ electrons/cm$^2$ and that of the bottom one is $1.7 \times 10^{11}$ electrons/cm$^2$. In the presence of high magnetic fields, each of the two layers forms its own ladder of spin-splitted Landau levels. The relative alignment of the two Landau-level ladders and the total electron density can be tuned in our system by an external electric field applied perpendicularly to the layers by means of a Schottky gate evaporated onto the top surface. When appropriate values of the magnetic field and gate voltage ($V_g$) are chosen, Landau levels with opposite pseudospin values, i.e. originating from the two different layers, can be brought close to alignment and at filling factors 4 and 2 ground states with Ising ferromagnetic order can be created[19]. A first-order phase transition that separates a ferromagnetic from a paramagnetic configuration was reported as a function of gate voltage and/or the magnetic field[19]. Additional evidence of this phenomenon in the capacitance signal is shown in Fig. 1.

$C_B$ measurements were performed by adding a small AC voltage (0.5 mV at 15.6 Hz) to the DC gate bias and monitoring the out-of-phase component of the current. The latter reflects the geometrical capacitance of a coupled-layer system and its density of states at the Fermi level in a perpendicular magnetic field provided that the frequency $f$ of the excitation signal fulfils the condition $f << \sigma_{xx}/C_B$, where $\sigma_{xx}$ is the bulk conductance. In the opposite limit, observed non-zero values of the capacitance signal indicate the presence of regions of the sample with finite area and conductance values much larger than $\sigma_{xx}$. With particular sample geometries this has allowed[18] to measure the edge channels widths at various filling factors.

### 3. Results

Figure 1A shows the measured $C_B$ for up and down sweeps of the magnetic field at $T$ = 30 mK and $V_g$ = -40 mV, close to a total filling factor $\nu$ = 4. At this $V_g$ the $\nu$ = 4 QH state is strongly suppressed and pronounced hysteresis is observed around its minimum. This behaviour is indicative of the development of out-of-equilibrium domain structures with long relaxation times compared to the experimental timescales. The suppression of the QH gap at the transition point is clearly visible in the capacitance signal as a function of both magnetic field B and gate voltage $V_g$, shown in Fig. 1C.

The suppression of the QH gap at the transition is clearly visible in the capacitance signal as a function of both $B$ and $V_g$, shown in Fig. 1C. "P" and "U" mark the polarized and un-polarized configurations respectively. Along the dashed line that connects points "P" and "U" at fixed $\nu$ = 4 the QH state is found to disappear at the crossing among opposite-pseudospin, opposite-spin Landau levels marking the occurrence of the phase transition. The Landau level spectrum in the "P" and "U" QH regions is sketched in Fig. 1B, where black and red lines indicate opposite-pseudospin levels and vertical arrows show the preferred spin orientation of the corresponding Landau levels. At the onset of this phase transition the two Landau level configurations are both self-consistent local-density approximation solutions[19]. Local fluctuations of the potential profile can thus break the system into domains of opposite magnetization in agreement with the observed magneto-transport behavior. Similar experimental observations characterized the crossing at $\nu$ = 2. At $\nu$ = 5, on the contrary, a clear anti-crossing behaviour near Landau level coincidence is found. Figure 2 displays resistivity data versus magnetic field and gate voltage. The anticrossing behaviour at $\nu$ = 5 and the gap collapse at $\nu$ = 4 are clearly visible. At $\nu$ = 5 the two levels at coincidence have the same (real) spin thus favouring the development of an easy-plane anisotropic state.

As shown in the following this hysteretical behavior originates from the out-of-equilibrium dynamics of metastable magnetic domains towards energy minima which persists during magnetic field sweeps. In order to investigate this relaxation we drove the system reproducibly in an out-of-equilibrium configuration and monitored its time evolution. To this end, we moved the system to a point of the $B$, $V_g$ plane (*reset*) far from the phase transition by applying an appropriate gate bias. After a delay $t_{delay}$, the gate bias was swept to the value of interest (*start*) and $C_B$ was monitored as a function of time. In our case, we set $t_{delay}$ = 120 s and the reset point was established near the $\nu$ = 5 ($V_g$ = 150 mV) QH state. The observed change of the capacitance signal during repeated cycles of relaxation strongly depends on the exact value of the filling factor of the reset state. For reset states close to $\nu$ = 5, maxima at deviations $\pm\delta\nu \sim 0.3$ were observed. These results reproduce the behavior of the spin-lattice relaxation times found at $\nu$ = 1 as well as at other integer QH states[20,21]. Together with the observed long relaxation times these data indicate the involvement of nuclei as the storage medium of the domain configuration and are consistent with recent experiments showing directly that these spin-flips induce simultaneous reversal of nuclear spins[22].

A representative example of the magneto-capacitance relaxation is shown in Fig. 3A. The capacitance is observed to rise steadily for about 1000 s and then evolve with jumps (same panel, right vertical axis) for $t$ > 1500 s. These features strongly resemble the Barkhausen jumps characteristic of conventional ferromagnetic materials and suggest that domains at longer time scale evolve by jumps between local energy minima separated by large energy barriers. In these experiments the growth of $C_B$ results from the increase of length of metastable magnetic domain loops, a process that requires electron-spin-flip events[23]. This out-of-equilibrium behaviour is concomitant with the creation of the Ising QH ferromagnet and is pervasive of the $V_g$, $B$ plane regions in the vicinity of Landau level crossing at $\nu$ = 4 where hysteresis is found. None of these features were observed away from the critical regions close to the crossing point. This pinpoints the link among slow relaxation, hysteresis and QH Ising ferromagnetism.

A careful investigation of the relaxation shows that it follows a stretched-exponential over more that two decades of time. This is demonstrated in Fig. 3B where the evolution shown in Fig. 3A is plotted together with the best fit with exponential and stretched-exponential ($C(t) = C_0 - A\exp[-(t/\tau)^\beta], 0 < \beta \leq 1$) trial functions. The stretched-exponential trial function gives a very accurate fit with $\beta = 0.35 \pm 0.05$ and $\tau = 72 \pm 15$ s. This non-exponential trend is observed in the entire phase-transition region as shown in Fig. 3C. This non-Arrhenius behaviour agrees with the idea of a metastable configuration of domains that evolves through a broad distribution of energy barriers. Figure 3C shows, in addition, that it is possible to tune *in situ* these non-equilibrium properties by changing the level alignment and the deviation of the filling factor from $\nu = 4$ with $B$ and $V_g$. As an example Fig. 4A reports the values of $\tau$ (left panel) and $\beta$ (right panel) at 30 mK as a function of $V_g$ and at three different magnetic field values. Above $V_g \approx -52$ mV, the 2DEG develops the $\nu = 4$ QH state and the amplitude of the relaxation becomes too small to be fitted. Close to this value the relaxation approaches an Arrhenius-like behaviour with $\beta$ coefficient up to 0.8. As we move to lower $V_g$s, the relaxation time $\tau$ first develops a peak (with values as high as 400 s) and then becomes vanishingly small below a critical value of $V_g^c \approx -62$ mV. Correspondingly, the values of the $\beta$ coefficient drop from about 0.8 to 0.3 signalling that disorder plays an increasingly important role in the relaxation process.

**Discussion**

In order to understand the physical implications of these observations, we recall that the energy of DW excitations determines both the current able to pass across a DW and the rate of relaxation through hyperfine interaction with the nuclei. It has been shown, in fact, that gapless Goldstone modes associated to spin rotation symmetry U(1) about the magnetic field direction can occur in the absence of spin-orbit effects[14,24,25]. With such low-energy modes, scattering across DWs is forbidden and current propagates along the walls determining high values of both resistance and capacitance[14]. In addition, recent calculations indicate that these soft spin modes allow fast spin relaxation ~ 0.1 s through hyperfine coupling with nuclei[24]. In the presence of spin-orbit coupling, the U(1) symmetry is broken and a small gap and a linear dispersion characterize the spectrum of DW excitations. An analytical expression of this gap is available only for the case $\nu = 1$ and without the Hartree interaction term[25]. In this case the magnetic domain structure can arise from local fluctuations of the Zeeman energy $\Delta_Z$ and the gap is proportional to $\Delta_Z$.[25] This situation is similar to the one here discussed once the Zeeman gap is replaced by the single-particle gap $\Delta$ that separates the opposite-spin, opposite-pseudospin Landau levels shown in Fig. 1B. Since $\Delta$ is reduced by lowering $V_g$ our system offers the possibility to emphasize the impact of soft DW excitations in the relaxation dynamics by tuning their energy. Indeed, the data presented in Fig. 4A support this theoretical framework: a decrease of the relaxation time is induced by lowering $V_g$ and correlates to the increase of the capacitance signal (see Figs. 3C and 4). The relaxation times found at low values of $V_g$ (below 1 s) are also in quantitative agreement with the calculated value of Ref. 24. Such values are much shorter than typical nuclear relaxation times (of the order of hundred of seconds) and confirm the role of soft spin waves in determining the coupling with nuclei and relaxation dynamics.

Figure 4B displays the relaxation time $\tau$ versus $V_g$ at four different temperatures with $B$ = 3.84 T. It can be noted that the critical gate voltages corresponding to $\tau \sim 0$ shift towards higher values when the temperature is increased indicating that a metastable phase of electrons exists only in an appropriate range of gate-voltage and low-temperature values (see the inset of Fig. 4B where the $V_g^c$ are plotted as a function of temperature). Data also suggest that gate bias and temperature play interchangeable roles in determining the properties of the system. To further highlight this point we show in Fig. 5 the evolution of hysteresis curves recorded at different $V_g$ (Fig. 5A) and at different temperatures (Fig. 5B). The remarkable similarities between the two set of data extend to the whole range of $V_g$ and $T$ where the metastable phase occurs. It is thus possible to create states characterized by the same dynamics at different temperatures by applying appropriate values of the gate voltage. This behaviour can be exploited to further investigate the properties of DWs between ferromagnetic QH states.

**Acknowledgments -** We thank R. Fazio, G.F. Giuliani, and A.H MacDonald for useful discussions. The work at NEST-INFM, Scuola Normale Superiore was supported in part by MIUR.

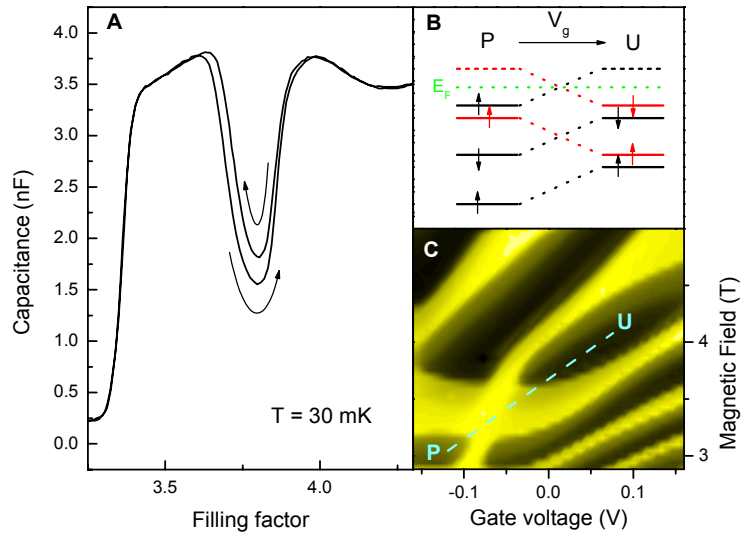

**Fig. 1** (A) Representative trace of the capacitance at $T$ = 30 mK for up and down sweeps of the magnetic field (indicated by arrows) at $V_g$ = -40 mV. (B) Schematic representation of the evolution of Landau energy levels as a function of $V_g$ at total filling factor $\nu$ = 4. Red (black) lines represent levels corresponding to states that occupy the bottom (top, closer to the surface) two-dimensional electron gas, pseudospin up and down, respectively. At $\nu$ = 4 a crossing occurs between spin-up pseudospin-down and spin-down pseudospin-up Landau levels which separates phases U and P characterized by two different spin and pseudospin configurations. Phase P is partially spin-polarized; phase U is spin un-polarized. (C) Plot of the capacitance as a function of magnetic field and gate voltage. Dark (bright) regions

represent low (high) values of the capacitance. The dotted line shows the evolution of the quantum Hall minimum at $\nu = 4$.

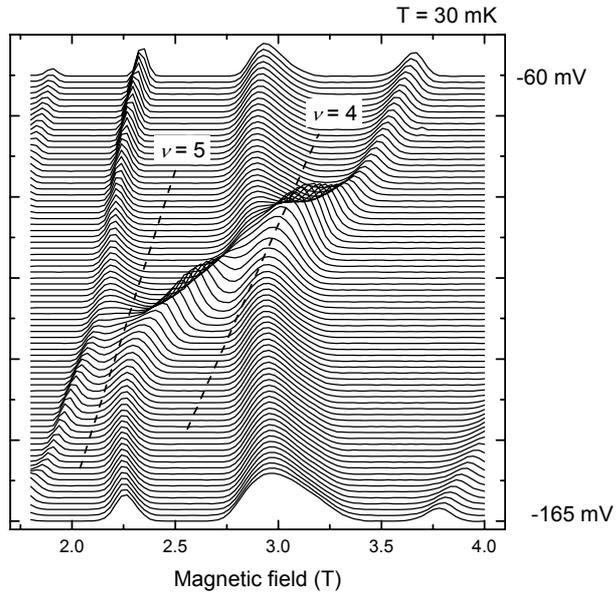

**Fig. 2** Resistivity data versus magnetic field and gate voltage at 30 mK. Curves corresponding to different gate voltages, ranging from to -165 mV (bottom curve) to -60 mV (top curve), are vertically offset for clarity. The dashed lines labels the evolution of the $\nu = 4$ and $\nu = 5$ states. The anticrossing behaviour at $\nu = 5$ and the gap collapse at $\nu = 4$ are clearly visible.

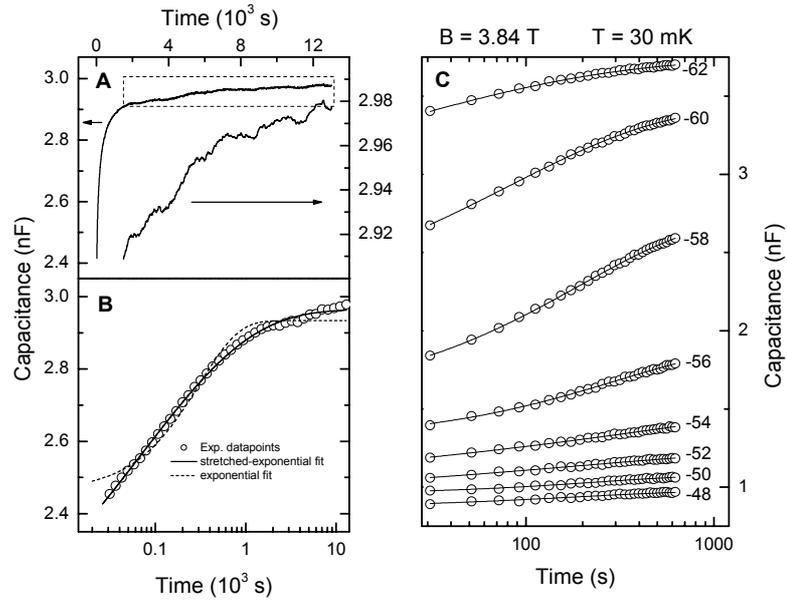

**Fig. 3** (A) Time evolution of the magneto-capacitance at $V_g$ = -65 mV, $B$ = 3.72 T and $T$ = 30 mK after excursion from $\nu$ = 5 ($V_g$ = 150 mV) to $V_g$ = -65 mV at fixed magnetic field. Data are shown after the initial 25 s in order to avoid fast equilibration effects after the gate voltage sweep. An enlarged view of the longer time scale behaviour (after 1500 s) is also shown (right vertical axis). In this region randomly located capacitance jumps are observed. Similar jumps can be detected at different $V_g$s close to the disappearing of the $\nu$ = 4 quantum Hall state. (B) Exponential and stretched exponential best fits ($\beta$ = 1, $\tau$ = 310±5 s and $\beta$ = 0.35±0.05, $\tau$ = 72±15 s, respectively) of the relaxation curve shown in (A). (C) Capacitance relaxation from $V_g$ = -48 mV to $V_g$ = -62 mV at $B$ = 3.84 T and $T$ = 30 mK. Solid lines are the results of the best-fit procedure with the stretched exponential function. Values for $\tau$ and $\beta$ are reported in Fig. 4.

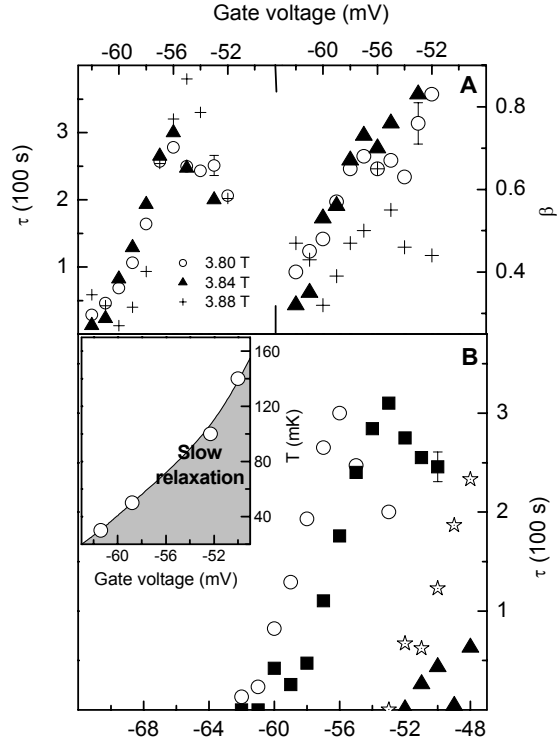

**Fig. 4** (A) Time constant $\tau$ (left axis) and stretching factor $\beta$ (right axis) as a function of gate voltage for three different magnetic fields at $T = 30$ mK. (B) Evolution of time constants $\tau$ versus gate voltage at temperatures of 30 mK (open dots), 50 mK (filled squares), 100 mK (stars), and 140 mK (filled triangles) with $B = 3.84$ T. (Inset) Phase diagram as a function of temperature and gate-voltage. At a fixed temperature the critical gate voltage value corresponds to $\tau = 0$ within our experimental uncertainty. The shaded region indicates the bias-temperature region in which slow relaxation is observed.

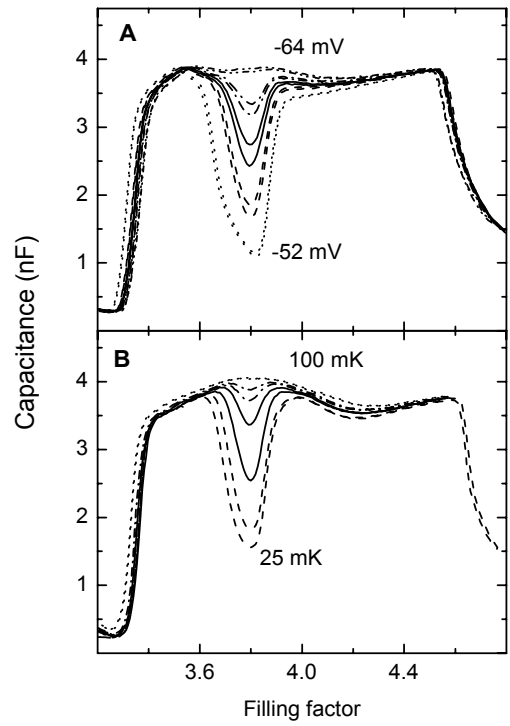

**Fig. 5** (A) Traces of the capacitance at $T = 30$ mK for up and down sweeps of the magnetic field at different values of gate voltage (from $V_g = -64$ mV to $V_g = -52$ mV). The sweep rate is 0.1 T/min. (B) Same as in (A) but at a fixed gate voltage $V_g = -40$ mV and at different temperatures (from $T = 25$ mK to $T = 100$ mK).